\documentclass[natbib]{emulateapj}

\def\s2n{S^{\prime}/N}

\begin{document}
\title{The Power Spectrum of Turbulence in NGC~1333: Outflows or Large-Scale Driving?}
\author{Paolo Padoan\altaffilmark{1}, Mika Juvela\altaffilmark{2},
Alexei Kritsuk\altaffilmark{1} and Michael L. Norman\altaffilmark{1}}
\altaffiltext{1}{Department of Physics, University of California, San Diego, 
CASS/UCSD 0424, 9500 Gilman Drive, La Jolla, CA 92093-0424; ppadoan@ucsd.edu}
\altaffiltext{2}{Department of Astronomy, University of Helsinki,
T\"ahtitorninm\"aki, P.O.Box 14,FI-00014 University of Helsinki, Finland}

\begin{abstract}

Is the turbulence in cluster-forming regions internally driven by stellar outflows or the consequence of a
large-scale turbulent cascade? We address this question by studying the turbulent energy spectrum in 
NGC~1333. Using synthetic $^{13}$CO maps computed with a snapshot of a supersonic turbulence simulation,
we show that the VCS method of Lazarian and Pogosyan provides an accurate estimate of the turbulent energy 
spectrum. We then apply this method to the $^{13}$CO map of NGC~1333 from the COMPLETE database. We find 
the turbulent energy spectrum is a power law, $E(k)\propto k^{-\beta}$, in the range of scales 0.06~pc $\le \ell \le 1.5$~pc, 
with slope $\beta=1.85\pm 0.04$. The estimated energy injection scale of stellar outflows in NGC~1333 is 
$\ell_{\rm inj}\approx 0.3$~pc, well resolved by the observations. There is no evidence of the flattening of the 
energy spectrum above the scale $\ell_{\rm inj}$ predicted by outflow-driven simulations and analytical models. 
The power spectrum of integrated intensity is also a nearly perfect power law in the range of scales 0.16~pc$<\ell<$7.9~pc, 
with no feature above $\ell_{\rm{inj}}$. We conclude that the observed turbulence in NGC~1333 does not appear to be 
driven primarily by stellar outflows.

\end{abstract}


\section{Introduction}

The structure of density and velocity fields in giant molecular clouds can be characterized by 
extended power laws (e.g. Heyer and Brunt 2004; Padoan et al. 2004, 2006), with scaling exponents 
consistent with those of numerical simulations of supersonic turbulence (e.g. Padoan et al. 2007; Kritsuk
et al. 2007, 2009a,b). 
The existence of these scaling laws and of one-point statistics of turbulent flows (for example the probability
distribution of gas density) is an important assumption in statistical theories of star formation aimed at the 
prediction of the stellar initial mass function (Padoan and Nordlund 2002, 2004; Hannebelle and Chabrier
2008, 2009) and the star formation rate (Krumholz and McKee 2005; Padoan and Nordlund 2009). 

Is this assumption of a universal large-scale turbulent cascade valid in cluster-forming regions, believed to 
be the birthplace of most stars, or is the turbulence there driven internally by stellar outflows? 
As an alternative to the idea that stellar clusters are formed in one crossing time (Elmegreen 
2000), Tan, Krumholz, and McKee (2006) propose that stars are formed in protocluster clumps over several 
dynamical times, while the turbulence is driven by stellar outflows. This scenario is investigated analytically 
by Matzner (2007) and numerically by Nakamura and Li (2007), Carroll et al. (2007), and Wang et al. (2009). 

Nakamura and Li (2007) find that the velocity power spectrum in their outflow-driven simulation is very shallow above
a characteristic scale of energy injection by outflows, $\ell_{\rm{inj}}$. They derive $\ell_{\rm inj}\approx0.3$~pc, 
assuming their computational domain has a size $L=1.5$~pc and a total mass $m_{\rm tot}=929$~m$_{\odot}$, 
characteristic of cluster-forming regions. This value of $\ell_{\rm inj}$ and the flattening of the power spectrum at larger 
scales is consistent with the analytical results in Matzner (2007). Carroll et al. (2009) confirm the results of Nakamura
and Li (2007) at higher numerical resolution, producing velocity power spectra even more clearly peaked at the 
outflow driving scale. With the values of mass, column density, and outflow 
momentum per unit mass, $v_{\rm c}=50$~km/s, adopted by Nakamura and Li (2007), equation (38) in Matzner (2007) 
gives $\ell_{\rm inj}=0.27$~pc. The injection scale is only weakly dependent on $v_{\rm c}$, $m_{\rm tot}$, and the star formation 
rate, and is approximately given by $\ell_{\rm inj}\approx0.079\,\Sigma_{\rm cgs}^{-15/28}$~pc, where $\Sigma_{\rm cgs}$ 
is the column density in g/cm$^2$. Our conclusions do not critically depend on a precise knowledge of $\ell_{\rm inj}$, 
as long as scales just above $\ell_{\rm inj}$ are resolved in the observations.

The prototype of outflow-driven cluster-forming regions chosen by Nakamura and Li (2007), Matzner (2007), and
Carroll et al. (2009) is NGC~1333 in Perseus. In this Letter, we derive the 
power spectrum of integrated intensity, $P_I(k)$, and velocity, $P_v(k)$, in NGC~1333, based on the data from the 
Five College Radio Astronomy Observatory (FCRAO) survey of the Perseus molecular cloud complex (Ridge et al. 2006), 
publicly available from the COMPLETE website. We show that the scale $\ell_{\rm inj}$ is resolved by the observations, 
and both power spectra, $P_I(k)$ and $P_v(k)$, are consistent with power laws extending to large scale, with the same
slope found in supersonic turbulence simulations and in the Perseus complex on larger scale (Padoan et al. 2006). 
The constant slopes of $P_v(k)$ and $P_I(k)$ above the scale $\ell_{\rm inj}$ suggest that outflows may not be the dominant driving mechanism.
In a recent paper appeared after the submission of this Letter, Brunt, Heyer, and Mac Low (2009) have reached the 
same conclusion for NGC~1333 and for other molecular cloud regions using a different method.

\section{Power Spectrum from the VCS Method}

\begin{figure}[ht]
\centering
\epsfxsize=9cm \epsfbox{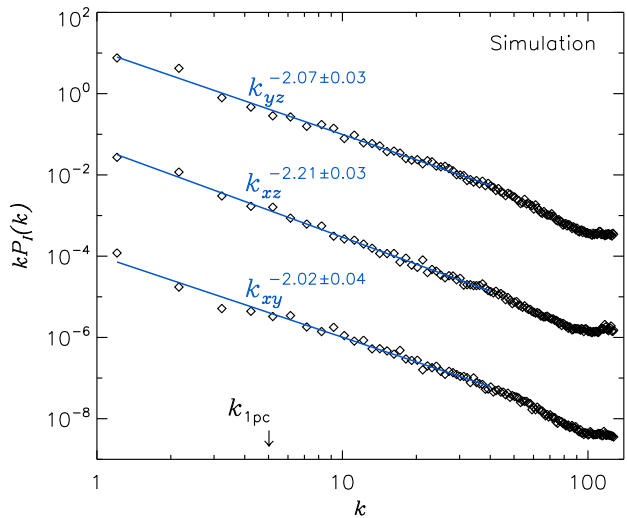}
\caption[]{Power spectra of integrated intensity, $P_I(k)$, from the three simulated $^{13}$CO maps. The wavenumber
corresponding to the scale of 1~pc is marked by an arrow. The plots are compensated by a factor of $k$, so 
the absolute values of the slopes given in the figure are the values of $\beta_v-1$ for the three maps.}
\label{f1}
\end{figure}

We first confirm the validity of the velocity coordinate spectrum (VCS) method of Lazarian and 
Pogosyan (2006), using synthetic maps of the $J$=1-0 line of $^{13}$CO, computed with a non-LTE 
radiative transfer code (Juvela and Padoan 2005), based on the density and velocity fields from a 
snapshot of a simulation of supersonic hydrodynamic turbulence with rms Mach number $\mathcal M$=6. 
These synthetic maps are the same used by Padoan et al. (2006) to test the VCA method of Lazarian and 
Pogosyan (2000) (Padoan et al. (2006) applied the VCA method to the full FCRAO Perseus map and 
found a power law turbulent energy spectrum, $E(k)\propto k^2P_v(k)\propto k^{-\beta}$, with the exponent 
$\beta=1.81\pm0.10$). 

The simulation is carried out with the {\em Enzo} code, developed at the Laboratory for 
Computational Astrophysics by Bryan, Norman and collaborators (Norman and Bryan 1999). 
{\em Enzo} is a public domain Eulerian grid-based code (see http://lca.ucsd.edu/projects/enzo) 
that adopts the Piecewise Parabolic Method (PPM) of Colella and Woodward (1984). We use an 
isothermal equation of state, periodic boundary conditions, initially uniform density and random 
large-scale velocity. The turbulence is forced in Fourier space only in the wavenumber range 
$1\le k\le2$, where $k=1$ corresponds to the size of the computational domain that contains 
$1,024^3$ computational zones (for details see Kritsuk et al. 2007). 

The radiative transfer calculations assume a box size of 5~pc, a mean density 
of $10^3$~cm$^{-3}$, a mean kinetic temperature of 10~K, an rms Mach number 
$\mathcal M$=6 (consistent with the turbulence simulation) and a uniform $^{13}$CO 
abundance of $10^{-6}$. These values were chosen as a generic reference model, not tailored 
to the Perseus molecular cloud complex. The density and velocity data cubes are resampled 
from $1,024^3$ to $256^3$ zones. The resampling of the data 
cubes has several advantages: i) It yields density and velocity fields with power spectra that 
are power laws almost up to the new Nyquist frequency; ii) It speeds up the radiative transfer 
calculations; iii) It generates a map of synthetic spectra with a range of scales comparable to that 
of the map of NGC~1333.

\begin{figure}[ht]
\centering
\epsfxsize=9cm \epsfbox{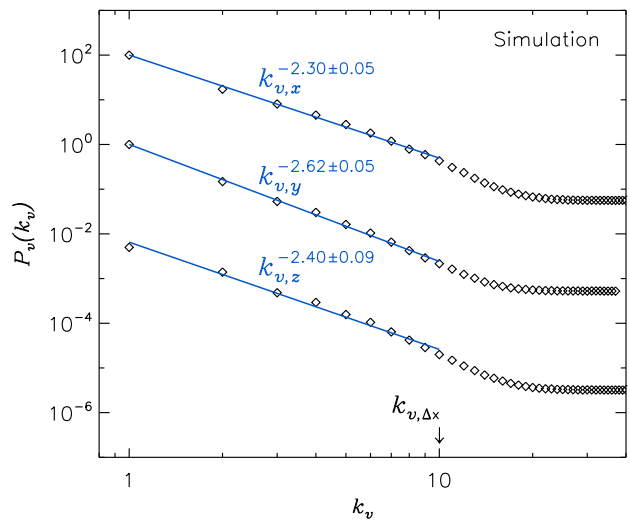}
\caption[]{Velocity coordinate power spectra from the three simulated $^{13}$CO maps. The wavenumber above
which the spectra are expected to steepen (in the absence of thermal broadening), $k_{v,\Delta x}$, is marked by 
an arrow. The absolute values of the least squares fit exponents given in the figure are the values of $\beta_v$ 
for the three maps.}
\label{f2}
\end{figure}

As a result of the radiative transfer calculations, we compute three maps (one in each of the three
axis directions) of spectral profiles of the $^{13}$CO ($J$=1-0) line intensity, $T({\bf x},v)$, where 
${\bf x}$ is the map position and $v$ the velocity channel. We compute $T({\bf x},v)$ for 280 velocity 
channels and $256^2$ map positions, so we obtain a position-position-velocity data cube of $256\times 256\times 280$ 
elements, for each of the three coordinate directions. The width of the velocity channels is
$\Delta v=0.025$~km/s, giving a total velocity range of 7~km/s across the 280 channels. The integrated 
intensity, $I({\bf x})$, is a two-dimensional map obtained as the sum of the line intensity over all the 
velocity channels, multiplied by the channel width, $I({\bf x})=\Sigma_vT({\bf x},v)\,\Delta v$. 

We call $P_I(k)$ the two-dimensional power spectrum of the integrated intensity map, $I({\bf x})$, 
and approximate it with a power law fit such that, $P_I(k)\propto k^{-\beta_I}$. We then call $E(k)$ 
the energy spectrum, given by $E(k)\propto k^2P_v(k)$, where $P_v(k)$ is the three
dimensional power spectrum of the velocity field, also assumed to be a power law, 
$E(k)\propto k^2P_v(k)\propto k^{-\beta}$. The VCS method of Lazarian and Pogosyan (2006) allows the 
determination of the energy spectrum (the exponent $\beta$), directly from the 
one dimensional power spectrum in velocity space, $P_1({\bf x},k_v)=|\int dv\, e^{i k_v  v}\, T({\bf x},v)|^2$,
that is the power spectrum of individual spectral line profiles. To reduce the effect of instrumental noise and of 
intrinsic variations in the power spectrum across the map, a global power spectrum, $P_v(k_v)$, is obtained as 
the sum of $P_1({\bf x},k_v)$ over a region of the map, $P_v(k_v)=\Sigma_{\bf x}P_1({\bf x},k_v)$.
If $\beta_v$ is the exponent of the velocity coordinate power spectrum, $P_v(k_v)\propto k_v^{-\beta_v}$, 
the turbulent energy spectrum exponent is given by $\beta=1+2/\beta_v$, if two conditions are satisfied:
1) a steep density power spectrum, $P_{\rho}(k)\propto k^{-\beta_{\rho}}$, with $\beta_{\rho}\ge3$, and 
2) small velocity wavenumbers, $k_v<k_{v,\Delta x}$, with $k_{v,\Delta x}=(\Delta x/S)^{-m/2}$, where 
$\Delta x$ is the observational beam size, $S$ is the size of the mapped region, and $m$ is the exponent of 
the second order velocity structure function, $m=\beta -1$. 

The first condition, $\beta_{\rho}\ge3$, is here approximated as $\beta_I \ge3$, because the power spectrum 
of the integrated intensity, $P_I(k)$, is expected to provide a good estimate of the density power spectrum. This 
is confirmed by our study of the synthetic spectral maps (see below), showing that $\beta=1+2/\beta_v$ while 
$\beta_I \ge3$. The second condition means that the exponent $\beta_v$ must be evaluated only for large velocity 
scales (small velocity wavenumbers) relative to the characteristic turbulent velocity at the scale of the beam size, 
$\Delta x$. We have made the velocity (hence $k_v$) nondimensional, $v\rightarrow v/(N_{\rm tot}\times \Delta v)$, 
where $N_{\rm tot}$ is the total number of velocity channels used to compute the power spectrum, and have defined the 
velocity wavenumber as $k_v=1/v$, so $k_v=1$ corresponds to the total velocity range used to compute the power 
spectrum. Because the characteristic turbulent velocity scales as $v(\ell)\propto\ell^{m/2}$, the velocity wavenumber 
scales as $k_v(\ell)\propto\ell^{-m/2}$, explaining the dependence of $k_{v,\Delta x}$ on the beam size, $\Delta x$.  

Before computing the power spectra, we apply a Gaussian beam with FWHM=$dx$, where $dx$ is the 
numerical mesh size, and add Gaussian noise to the synthetic spectra, 
to a level comparable to that of the FCRAO map of NGC~1333.
The power spectra $P_I(k)$ and $P_v(k_v)$ are shown for all three directions in Figures~1 and 2 respectively. $P_I(k)$ is very
well described by a power law in the approximate range $1<k<50$ ($k=1$ corresponds to the linear 
size of the computational domain). Averaging the results of the three maps, we obtain the exponent $\beta_I=3.10\pm 0.08$,
where the uncertainty is the standard deviation of the three measurements, a little larger than the standard deviation
of each of the three individual least squares fits. In Padoan et al. (2006) we measured the same power 
spectrum, but without adding noise to the synthetic maps, and only in the direction of the x-axis. We obtained 
a slope of $2.99\pm0.08$, consistent with the slope of $3.07$ found here for the same direction. 
This slope is also consistent with that of the density power spectrum measured directly from the three-dimensional 
snapshot, $\beta_{\rho}=3.0$. 

The least squares fits to the $P_v(k_v)$ spectra are computed in the range $1<k_v<k_{v,\Delta x}$, 
where $k_{v,\Delta x}=256^{(\beta-1)/2}=10$, to satisfy the second condition for the validity of the VCS method. 
As shown in Figure~2, the power spectra $P_v(k_v)$ are power laws up to $k_v\approx k_{v,\Delta x}$, and become 
steeper at larger values of $k_v$, as predicted by the theory. The steepening is here reduced by the effect of the 
added noise.
Because the condition $\beta_I \ge3$ is satisfied, we derive the energy spectrum from the relation 
$\beta=1+2/\beta_v$. We obtain $\beta=1.82\pm0.04$, where the uncertainty is the standard deviation 
\begin{figure}[ht]
\centering
\epsfxsize=9cm \epsfbox{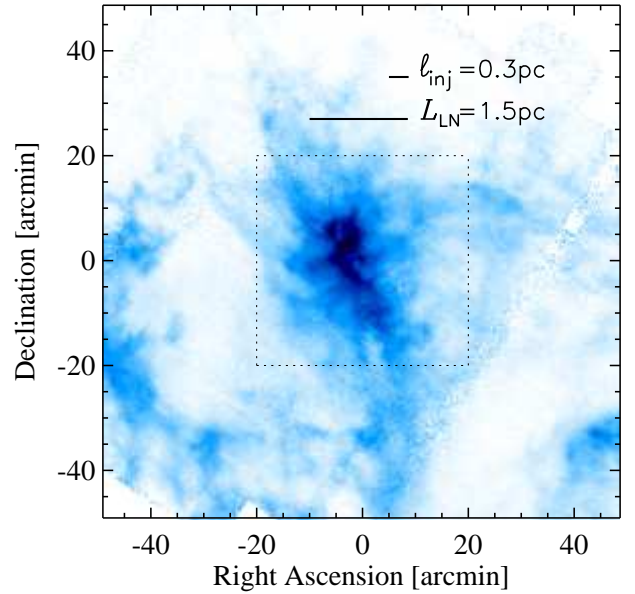}
\caption[]{Integrated intensity map of NGC~1333, from the region of the COMPLETE survey of Ridge et al. (2006) used in
this work. The power spectra are also computed within the smaller square region marked by dotted lines, with essentially 
the same result. The two segments indicate the size of the outflow injection scale, $\ell_{\rm inj}$, and the computational box size, 
$L_{\rm LN}$, in the simulation of Nakamura and Li (2007), assuming a distance to NGC~1333 of 280~pc. Assuming this distance,
the map size is 7.9$\times$7.9~pc, while the smaller map is 3.2$\times$3.2~pc.}
\label{f3}
\end{figure}
of the exponents from the three maps. This result is consistent with the slope of 1.8 derived directly from the three-dimensional 
velocity field of the simulation snapshot. The value of $\beta_I$ is very close to the value of 3 that separates the 
regimes of steep and shallow density power spectra discussed in 
Lazarian and Pogosyan (2000, 2006). This is not a problem, because in the case of shallow density power spectra, 
the relation between $\beta$ and $\beta_v$
would be either the same or $\beta=1+2(\beta_{\rho}-2)/\beta_v$, depending on some other condition. The
two formulas give the same result for $\beta_{\rho} \rightarrow 3$.

We conclude that the VCS method provides a precise estimate 
of the turbulent energy spectrum, by fitting the slope of  $P_v(k_v)$ for $k_v<k_{v,\Delta x}$. The $k_v$ range of the power 
law is limited. It is only a factor of 10 for the synthetic maps (and only a factor of 4 for the map of NGC~1333), because
it is essentially given by the spatial dynamical range (256 for the synthetic maps) to the power $m/2=(\beta-1)/2$, 
as explained above. However, the average of the power spectra over many positions on the map ($256^2$ for both the 
synthetic maps and the map of NGC~1333), yields a $P_v(k_v)$ spectrum that is extremely smooth and very well approximated
by a power law, providing an accurate estimate of the slope of the turbulent energy spectrum.

\section{The Power Spectra of NGC~1333}

We now apply the VCS method to the J=1-0 $^{13}$CO survey of the Perseus molecular cloud complex 
carried out with the FCRAO 14~m antenna by Ridge et al. (2006). The grid spacing of the survey is 23'', 
and the beam size 46''. The velocity-channel size is 0.06~km/s. We select a squared region of 
\begin{figure}[ht]
\centering
\epsfxsize=9cm \epsfbox{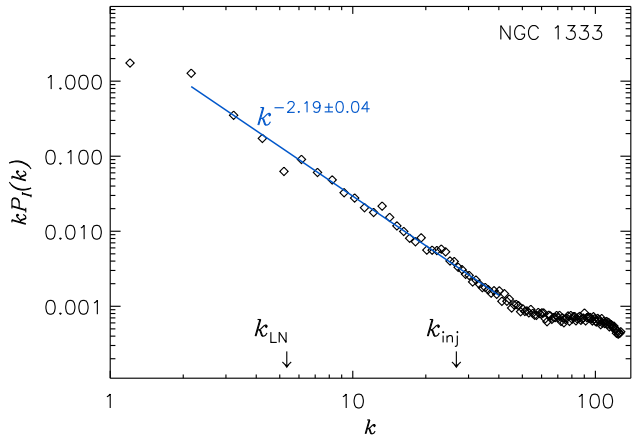}
\caption[]{Power spectrum of the integrated intensity map of NGC~1333 shown in Figure~\ref{f3}, compensated by 
a factor $k$, $k P_I(k)$. The wavenumbers corresponding to the outflow injection scale, $k_{\rm inj}$, and the 
computational box size, $k_{\rm LN}$, in the simulations of Nakamura and Li (2007) and Carroll et al. (2009), 
assuming a distance to 
NGC~1333 of 280~pc, are marked by two arrows. The wavenumber $k=1$ corresponds to the map size of 7.9~pc. 
The power spectrum is well appoximated by a power law in the range $1< k < 50$, and the least squares fit is 
computed in the range $2 \le k \le 40$.}
\label{f4}
\end{figure}
100$\times$100 arcmin, approximately centered on NGC~1333. This region contains $256\times 256$ spectra, 
like the synthetic maps.
Figure~3 shows the integrated intensity map of that region. It also shows the size of $\ell_{\rm inj}=0.3$~pc and
$L_{\rm LN}=1.5$~pc, the outflow injection scale and the size of the computational domain in Nakamura and Li (2007)
and Carroll et al. (2009),
assuming a distance to NGC~1333 of $D=280$~pc. For this distance, the size of the map is $S=7.9$~pc, the grid spacing
0.03~pc, and the beam size $\Delta x=0.06$~pc. The injection scale is therefore well resolved. The uncertainty on the distance 
to NGC~1333 is not very large, 220~pc$<D<$350~pc (Borgman and Blaauw 1964; Herbig and Jones 1983; Cernis 1990). 
Even assuming the largest value of $D=350$~pc, $\ell_{\rm inj}$ would still be 4 times larger than the beam size.

We compute the power spectra $P_I(k)$ and $P_v(k_v)$ within that region, without any correction for the 
effects of beam and noise. This is justified because we do not compute the least squares fit at very large 
values of $k_v$ and $k$, and because we have confirmed the validity of the VCS method 
using synthetic maps where noise was added to a comparable level as in the map of NGC~1333.
The $P_I(k)$ power spectrum is shown in Figure~\ref{f4}. The least squares fit, 
computed in the range $2\le k \le 40$, gives $\beta_I=3.19\pm 0.04$. The values of the wavenumbers 
$k_{\rm inj}$ and $k_{\rm LN}$, corresponding to the scales $\ell_{\rm inj}=0.3$~pc and
$L_{\rm LN}=1.5$~pc, are marked by two arrows. The power spectrum is an almost perfect power law 
in the range of wavenumbers $1\le k \le 50$, corresponding to the range of scales 0.16~pc$<\ell<$7.9~pc 
($k=1$ corresponds to the size of the map), with no significant feature around $\ell_{\rm inj}$.

\begin{figure}[ht]
\centering
\epsfxsize=9cm \epsfbox{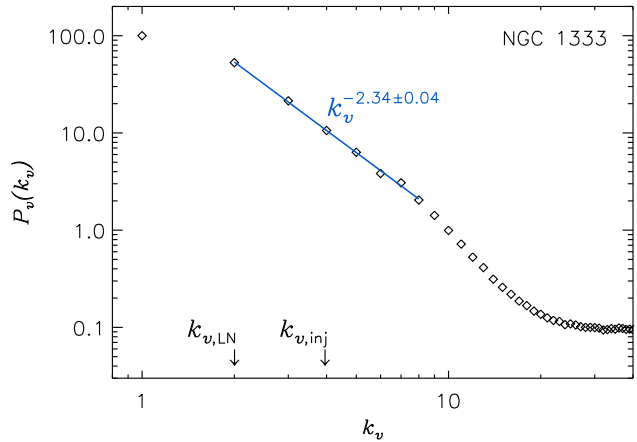}
\caption[]{Velocity coordinate power spectrum from the PPV data cube corresponding to the map of NGC~1333 shown
in Figure~\ref{f3}. The velocity wavenumbers corresponding to the outflow injection scale, $k_{v,\rm inj}$, and the 
computational box size, $k_{v,\rm LN}$, in the simulations of Nakamura and Li (2007) and Carroll et al. (2009), 
assuming a distance to NGC~1333 of 280~pc, are marked by two arrows. The dynamical range in velocity 
coordinates is compressed, relative to the range in spatial scales, as explained in the text. The least squares fit, computed 
in the range $2\le{k}_v\le8$, gives the value of $\beta_v=2.3\pm0.04$, used in the VCS method to derive the exponent of 
the turbulent energy spectrum, $\beta=1.85\pm0.02$.}
\label{f5}
\end{figure}

Figure~\ref{f5} shows the velocity coordinate power spectrum, $P_v(k_v)$, and 
its least squares fit in the range $2\le k_v\le8$, giving $\beta_v=2.34\pm0.04$. This is 
the range of $k_v$ values expected from the VCS theory, because here $k_{v, \Delta x}=128^{(\beta-1)/2}=7.7$.
As explained above, the dynamical range in $k_v$ is compressed with respect to that in $k$, as shown by the 
positions of $k_{v,\rm inj}$ and $k_{v,\rm LN}$, marked by two arrows. However, the measurement of the slope 
is still very accurate because the power spectrum is extremely smooth, as the result of averaging over the 
whole map, and very well represented by a power law. The injection scale falls in the very middle of the power 
law range, at $k_{v,\rm inj}=3.95$, and there is no sign of a variation of the slope within a range of values of $k_v$
corresponding to $\Delta x\le\ell\le L_{\rm LN}$, or 0.06~pc $\le\ell\le1.5$~pc. 

Using the relation $\beta=1+2/\beta_v$, we obtain $\beta=1.85\pm0.02$, where the standard deviation is now
derived directly by that of the least squares fit for $\beta_v$. This uncertainty may seem unrealistically small.
Using the synthetic maps we have derived a standard deviation of 0.04 from the three values of $\beta$. If we
instead derive the uncertainty directly from the standard deviation of each fit of $\beta_v$, we find that the 
average of the three uncertainty is 0.02, exactly like in the observational map. We therefore estimate that a more
realistic uncertainty for the derived slope in NGC~1333 should be $\approx0.04$. 

To further verify this result, we also select a smaller region of $40\times40$~arcmin ($3.2\times3.2$~pc) around 
NGC~1333, shown as a dotted square in Figure~\ref{f3}. We obtain $\beta_v=2.29\pm0.09$, corresponding
to $\beta=1.87\pm0.03$, consistent with the value estimated on the larger scale, though with a slightly larger
uncertainty due to the smaller dynamical range and the smaller number of $P_v(k_v)$ spectra in the map.
The good consistency between the two maps is not surprising, because the sum of the $P_v(k_v)$ spectra is 
dominated by the positions with the strongest integrated intensity that are mostly located within the inner map. 

We conclude that the VCS method yields a turbulent energy power spectrum with slope $\beta=1.85\pm0.04$, 
in the range of scales 0.06~pc $\le\ell\le1.5$~pc, with no evidence of internal driving by outflows at the scale
$\ell_{\rm inj}=0.3$~pc.

\section{Conclusions}

It has been suggested that protocluster clumps actively form stars for several dynamical times, while being supported
by the turbulence internally generated by stellar outflows (e.g. Tan, Krumholz, and McKee 2006). This scenario is 
simulated by Nakamura and Li (2007) and Carroll et al. (2009), with physical parameters characteristic of NGC~1333. 
They find an energy injection
scale, $\ell_{\rm inj}\approx0.3$~pc, above which the turbulent energy spectrum is very shallow, up to the size of 1.5~pc 
of their computational domain. This value of $\ell_{\rm inj}$ and the flattening of the power spectrum above that scale are 
consistent with the results of the analytical model of Matzner (2007). If the turbulence in cluster-forming regions is primarily 
driven by outflows, the turbulent energy spectrum should flatten above the scale $\ell_{\rm inj}$, up to a scale where the 
external turbulent cascade dominates again. 

We have searched for such a feature in the energy spectrum of NGC~1333, the prototype of outflow-driven regions in the cited 
theoretical works. To this aim, we have applied the VCS method of Lazarian and Pogosyan (2006) to the $^{13}$CO
FCRAO map of NGC~1333, after successfully testing the method on simulated data. We have found the energy spectrum
is a power law, $E(k)\propto k^2P_v(k)\propto k^{-\beta}$, with $\beta=1.85\pm0.04$, in the range of scales 
0.06~pc $\le \ell \le 1.5$~pc, with no evidence of internal driving by outflows at $\ell_{\rm inj}=0.3$~pc. The power
spectrum of integrated intensity is also a power law, $P_I(k)\propto k^{-\beta_I}$, with $\beta_I=3.19\pm0.04$, in the range 
of scales 0.16~pc$<\ell<$7.9~pc, with no significant features above the predicted injection scale. 

These power spectra are consistent with those of large-scale driven simulations of supersonic turbulence, and with 
those measured on larger scale in Perseus  and other molecular cloud complexes. Although outflows from young 
stars are present in NGC~1333, the large-scale turbulent cascade appears to be the main energy source. The 
turbulence in NGC~1333 is either currently driven by significant mass inflow from larger scales, or in the process 
of being dissipated, until the time when winds, outflows, and ionizing radiation from stars will completely disperse 
the star-forming gas. \\

\acknowledgements

We utilized computing resources provided by the San Diego Supercomputer Center and by the National Center 
for Supercomputing Applications. M.J. was supported by the Academy of Finland Grants no. 105623 and 124620.
A.K. was supported in part by the National Science Foundation through grant AST-0607675.


\begin{thebibliography}

\bibitem[Borgman \& Blaauw(1964)]{1964BAN....17..358B} Borgman, J., \& Blaauw, A.\ 1964, \bain, 17, 358 
\bibitem[Brunt et al.(2009)]{2009A&A...504..883B} Brunt, C.~M., Heyer, M.~H., \& Mac Low, M.-M.\ 2009, \aap, 504, 883
\bibitem[Carroll et al.(2009)]{2009ApJ...695.1376C} Carroll, J.~J., Frank, A., Blackman, E.~G., Cunningham, A.~J.,
\& Quillen, A.~C.\ 2009, \apj, 695, 1376
\bibitem[Cernis(1990)]{1990Ap&SS.166..315C} Cernis, K.\ 1990, \apss, 166, 315 
\bibitem[Colella \& Woodward(1984)]{1984JCoPh..54..174C} Colella, P., \& 
Woodward, P.~R.\ 1984, Journal of Computational Physics, 54, 174 
\bibitem[Heyer \& Brunt(2004)]{2004ApJ...615L..45H} Heyer, M.~H., \& Brunt, C.~M.\ 2004, \apjl, 615, L45
\bibitem[Herbig \& Jones(1983)]{1983AJ.....88.1040H} Herbig, G.~H., \& Jones, B.~F.\ 1983, \aj, 88, 1040
\bibitem[Juvela \& Padoan(2005)]{2005ApJ...618..744J} Juvela, M., \& Padoan, P.\ 2005, \apj, 618, 744
\bibitem[Kritsuk et al.(2007)]{2007ApJ...665..416K} Kritsuk, A.~G., Norman, 
M.~L., Padoan, P., \& Wagner, R.\ 2007, \apj, 665, 416 
\bibitem[Kritsuk et al.(2009)]{2009ASPC..406...15K} Kritsuk, A.~G., Ustyugov, S.~D., Norman, M.~L., 
\& Padoan, P.\ 2009, Astronomical Society of the Pacific Conference Series, 406, 15
\bibitem[Kritsuk et al.(2009)]{2009JPhCS.180a2020K} Kritsuk, A.~G., 
Ustyugov, S.~D., Norman, M.~L., \& Padoan, P.\ 2009, Journal of Physics Conference Series, 180, 012020
\bibitem[Krumholz \& McKee(2005)]{2005ApJ...630..250K} Krumholz, M.~R., \& McKee, C.~F.\ 2005, \apj, 630, 250 
\bibitem[Lazarian \& Pogosyan(2000)]{2000ApJ...537..720L} Lazarian, A., \& 
Pogosyan, D.\ 2000, \apj, 537, 720 
\bibitem[Lazarian \& Pogosyan(2006)]{2006ApJ...652.1348L} Lazarian, A., \& Pogosyan, D.\ 2006, \apj, 652, 1348 
\bibitem[Matzner(2007)]{2007ApJ...659.1394M} Matzner, C.~D.\ 2007, \apj, 659, 1394 
\bibitem[Nakamura \& Li(2007)]{2007ApJ...662..395N} Nakamura, F., \& Li, Z.-Y.\ 2007, \apj, 662, 395
\bibitem[Norman \& Bryan(1999)]{1999numa.conf...19N} Norman, M.~L., \& Bryan, G.~L.\ 1999, 
ASSL Vol.~240: Numerical Astrophysics, 19 
\bibitem[Padoan \& Nordlund(2002)]{2002ApJ...576..870P} Padoan, P., \& Nordlund, {\AA}.\ 2002, \apj, 576, 870 
\bibitem[Padoan \& Nordlund(2004)]{2004ApJ...617..559P} Padoan, P., \& Nordlund, {\AA}.\ 2004, \apj, 617, 559 
\bibitem[Padoan et al.(2004)]{2004ApJ...604L..49P} Padoan, P., Jimenez, R., 
Juvela, M., \& Nordlund, {\AA}.\ 2004a, \apjl, 604, L49 
\bibitem[Padoan et al.(2006)]{2006ApJ...653L.125P} Padoan, P., Juvela, M., 
Kritsuk, A., \& Norman, M.~L.\ 2006, \apjl, 653, L125 
\bibitem[Ridge et al.(2006)]{2006AJ....131.2921R} Ridge, N.~A., et al.\ 
2006, \aj, 131, 2921 
\bibitem[Tan et al.(2006)]{2006ApJ...641L.121T} Tan, J.~C., Krumholz, 
M.~R., \& McKee, C.~F.\ 2006, \apjl, 641, L121 

\end{thebibliography}
\end{document}